# Energy Spectra of Secondary Particles Induced by Solar Energetic Proton Events and Magnetospheric Effects

Ashot Chilingarian, Balabek Sargsyan, Mary Zazyan
Yerevan Physics Institute

## Abstract

We investigate the energy spectra of secondary cosmic ray particles associated with two distinct solar events: the magnetospheric effect (ME) of 5 November 2023 and ground-level enhancement (GLE #74) of 11 May 2024. Using data from the SEVAN and Neutron Monitor networks and energy release histograms from particle spectrometers, we reconstruct spectra and identify key differences between ME and GLE. CORSIKA-based simulations reveal that MEs are caused by galactic protons below geomagnetic cutoff rigidities (Rc ≈ 7.1 GV at Aragats) penetrating the magnetosphere during geomagnetic storms, leading to localized flux enhancements at mountain altitudes but not at sea level. In contrast, SEP events initiated by GLEs can involve high-energy solar protons (>10 GeV), producing secondaries that reach sea level at middle latitudes. We present integral energy spectra and spatial correlation of detector responses, demonstrating that SEVAN's energy-resolved data offer new diagnostic tools for identifying hard-spectrum SERs. Our results refine the definition of ME and suggest a strategy for early warning of hazardous solar particle events based on real-time ground-based observations.

## Introduction

The sun influences the Earth through electromagnetic radiation, plasma, and high-energy particles. Although these particles have a small fraction of the energy of visible light, they provide insights into fundamental particle acceleration processes (Dorman, 1974). Additionally, they can offer timely information about solar explosions that may affect technologies in space and on the Earth's surface (Hapgood, 2011).

For billions of years, the Earth has faced bombardment from protons and fully stripped ions that are accelerated throughout the galaxy. The flux of these particles can fluctuate as the Sun travels through different galactic arms in its orbit around the galaxy's center. Additionally, explosions of nearby stars and changes in the geomagnetic field (GMF) can affect particle flux. Nevertheless, the GeV galactic cosmic ray (GCR) flux remains relatively constant over millions of years. In interactions with the atmosphere, GCRs initiate cascades of lower energy particles, known as extensive air showers (EASs, Auger et al., 1939). The majority of secondary particles are absorbed in the atmosphere. Some reach the Earth's surface and constitute a more-or-less stable flux of different species of cosmic rays, gamma rays, neutrons, electrons, muons, and others. This stable flux is modulated by the sun's activity. The Sun exhibits high variability, capable of changing radiation and particle flux intensities near Earth by several orders of magnitude in mere minutes. These fluxes significantly impact Earth, influencing climate, safety, and space and surface technologies. The Sun modulates the low-energy GCR in various ways. Its most energetic flaring processes can release energy up to $10^{33}$ erg within a few minutes (Cliver and Dietrich 2013).

In addition to broadband electromagnetic radiation, the explosive flaring process typically leads to the ejection of substantial amounts of solar plasma (coronal mass ejection – CME, Cane and Lario, 2006) and accelerates many electrons and ions (solar energetic particle event – SEP, Reames, 2017). Particles may be generated directly at the coronal flare site, allowing them to escape into interplanetary space, or they may be accelerated in shocks associated with CMEs that travel through the corona and interplanetary medium (Lockwood et al., 1990). Charged particles and neutral particles produced from collisions with dense solar plasma form what are known as solar cosmic rays (SCR, Dorman, and Venkatesan, 1993). Upon reaching Earth's vicinity, these particles are detected by particle spectrometers aboard satellites and space stations. The highest energy solar particles create air showers that can reach the Earth's surface, resulting in ground-level enhancements (GLEs, Poluianov et al., 2017).

Only a few SEPs provide enough energetic particles to be detected by surface particle detectors; so far, only 75 GLEs have been recognized over nearly 80 years. The latitudinal dependence of the GMF enables distributed particle detectors to register GCRs within the specific energy range for each location defined by geomagnetic cutoff rigidities (Smart and Shea, 2009). Depending on their location, particle detectors are sensitive to different parts of the SEP spectra. Charged particles can enter the atmosphere with energies exceeding a minimum value specific to geographical coordinates. The GMF redirects particles with lower energies to lower latitudes or into open space. The attenuation of particle cascades in the atmosphere also establishes a threshold for the minimum energy of the primary particles, noticeable at geographical coordinates where geomagnetic cutoff rigidity is at its lowest. The minimum energy of incoming cosmic rays in polar regions is determined by atmospheric rigidity, ranging from approximately 300 MeV at high altitudes in Antarctica to 400 MeV for sea-level detectors in the same area (Poluianov and Batalla, 2022). The highest energy threshold of 17 GeV is associated with equatorial sites at the maximum of the GMF. In practice, the minimum energies of registered particles are impacted by particle type, detector efficiency, data acquisition electronics, and the matter above the detector. Correcting the registered count rates according to the detector response function makes it possible to recover energy spectra within the accessible energy range.

Cosmic rays are monitored on Earth by several networks of particle detectors that measure secondary cosmic rays produced when primary protons and stripped nuclei interact with atmospheric atoms. These networks include the Neutron Monitor network (NM, Simpson, 1957; Mishev and Usoskin, 2020), the Solar Neutron Telescope (SNT) network (Muraki et al., 1995), the SEVAN network (Chilingarian et al., 2009), which also captures electron, gamma-ray, and muon fluxes, Spaceship Earth (Kuwabara et al., 2006), and the Global Muon Detector Network (GMDN, Munakata et al., 2000). The network of NMs covers almost the entire globe, from the Antarctic to the near-Arctic regions; flux intensities measured at different sites are accessible from the NM database (NMDB, Mavromichalaki et al., 2011). SEVAN network detectors are located atop mountains in Armenia, Germany, and Eastern Europe, while other networks are situated at various latitudes and longitudes. NM and SEVAN networks detect solar events surprisingly coherently (Karapetyan et al. 2024).

The GCR proton differential energy spectrum at 10 – 1000 GeV exhibits several features and can be approximated by the power law $dJ_p/dE \sim E^{2.6}$, where γ is between 2.6 and 2.8 (Abe et al., 2016; Yoon et al., 2017). The SCR flux at GeV energies is very weak, typically not reaching GeV levels; it is only during rare solar events, such as those in 1956 and 2005, that energy spectra of SCR were recorded by the NM network up to several GeV (see, for example, Fig. 4 of Bieber et al., 2013). By enhancing the high-energy muon flux (>5 GeV), researchers demonstrated the existence of solar protons with energies surpassing 20 GeV during the 20 January 2005 SEP (Bostanjyan et al., 2007; Chilingarian, 2009). The early identification of 'hard' spectra containing GeV protons is crucial, as many mid-level energy particles (50-100 MeV) arriving within minutes can cause significant damage to satellite electronics. An alternative method for recovering energy spectra involves measuring different particle types at the same geographical coordinates, such as neutrons and muons (Blanco et al., 2024) or neutrons, muons, electrons, and gamma rays (Chilingarian et al., 2005) or by utilizing particle intensities recorded at the same location but at various altitudes (Chilingarian and Reymers, 2007). In this way, using intensities measured by particle detectors on the slopes of Mt. Aragats, the spectral index of the 20 January 2005 GLE was estimated (Zazyan and Chilingarian, 2009). This same paper also explored the relationship between the measured secondary particle energy spectra and the primary SCR energy spectra. The most probable energies of GeV primary protons, which initiate particle showers producing MeV secondary electrons, gamma rays, neutrons, and muons, were estimated.

Surface particle detectors quantify the number of secondary particles striking their surfaces in a specified time frame. Measurements taken over one-second or one-minute intervals provide key insights into solar modulation's physical effects. However, it is impossible to differentiate between solar cosmic rays (SCR) and galactic cosmic rays (GCR) based solely on the detected secondary particles. The impacts of solar modulation appear as non-random variations in the detector count rate time series. The abundance of low-energy primaries at high latitudes can lead to GLEs of up to thousands of percent (Shea and Smart, 2004). Conversely, the boosts due to SCR are much less pronounced at mid to low latitudes, typically not surpassing 2-5%. For energies above 10 GeV, the GCR intensity increasingly surpasses that of the largest known solar energetic particle (SEP) events (see Fig. 1 in Chilingarian and Reimers, 2007), posing a significant challenge to detect the small SCR signal amidst the substantial GCR background. Low-statistics can sometimes produce seemingly false peaks. Several strategies are discussed to mitigate potential errors in detecting signal existence (Chilingarian and Hovsepyan, 2023).

Other solar modulation effects influence the intensity of GCR. The solar wind 'blows out' the lowest energy GCR from the solar system, altering the GCR flux intensity based on the solar cycle year. Large magnetized plasma clouds and shocks initiated by the interplanetary CME (ICME) travel at velocities up to 2000 km/s, disrupting the interplanetary magnetic field (IMF). Upon reaching the magnetosphere, they introduce anisotropy in the GCR flux near Earth, creating depletion regions that result in an anisotropic distribution of GCR. The size of the southward component of the IMF ($B_z$) correlates with the modulation effects that ICME has on the ambient population of galactic cosmic rays. When reaching the magnetosphere, the overall depletion of GCR triggered by interplanetary shocks and plasma clouds manifests as a decrease in the secondary cosmic rays detected by networks of particle detectors on the Earth's surface (Forbush decrease; Forbush, 1949; Usoskin et al., 2008). The relative decline in the count rate at the particle monitors is pronounced at high latitudes. Due to low magnetic cutoff rigidity, the primary protons and ions, which account for most of the detector's count rate, possess significantly low energy (less than a few GeV). Therefore, they are considerably depleted by disturbances in the IMF. The count rates of particle monitors at middle to low latitudes arise from primaries with energies much higher than 5 GeV. Consequently, the relative depletion at higher energies, and thus the reduction in the count rates of low-latitude monitors, will be less than those of high-latitude.

In turn, geomagnetic storms (GMS, Chilingarian, and Bostanjyan, 2010), which manifest as sudden changes in the Earth's magnetic field, can enhance the count rates of middle and low-latitude particle detectors without significantly affecting the count rates of high-latitude detectors (Dvornikov et al., 1988). When the magnetic field of an ICME is directed southward, it reduces the effective cutoff rigidity, resulting in the so-called magnetospheric effect (ME, Kudela et al., 2008). The abrupt change in the geomagnetic field, known as sudden storm commencement (SSC, Smith et al., 2021), occurs when the shockwave from fast solar wind reaches Earth's magnetosphere, contributing to the main decrease phase supported by the southward magnetic field. Thus, GCRs of lower energies, which are typically blocked by the magnetosphere, penetrate the atmosphere and generate additional secondary particles, increasing the count rate of monitors located at middle and low latitudes by 2-5%. At high latitudes, especially in Antarctica, where cutoff rigidity is very low, the count rates of particle detectors are primarily determined by the atmospheric cutoff, which is independent of GMS.

During the 23rd solar activity cycle, the Aragats Space Environmental Center (ASEC, Chilingarian et al., 2003) lacked sufficient spectrometers and spectra recovery methods. As solar activity increased towards the peak of the 25th cycle, ASEC was equipped with various spectrometers (Chilingarian et al., 2022) and appropriate strategies for addressing the inverse problem of CR (Chilingarian et al., 2024a), i.e., using the registered energy spectra of secondaries to identify the particle type and energy that causes a particle cascade in the atmosphere. These occurrences include FD followed by a GLE in the FD recovery stage, which have been discussed in numerous publications (Hayakava et al., 2024; Abunina et al., 2024; Diaz, 2024; Ram et al., 2024; Schennetten et al., 2024). This paper focuses on particle flux analysis, utilizing developed spectra recovery techniques to characterize November 5, 2023's magnetospheric effect and the solar events of May 10-11, 2024. Additionally, we aim to establish a refined definition of ME to differentiate the magnetospheric effect from other solar events effectively.

1. **Magnetospheric event on 5 November 2023**

The first feature of the magnetospheric effect is the absence of enhancements in data from Antarctic NMs. Figure 1 shows a time series from several high-latitude NMs demonstrating no enhancement. Only the DRBS monitor shows apparent enhancement. Still, it is not in Antarctica but the Geophysical Center of Dourbes (Belgium); see also Fig. 1 of Gil et al. (2024).

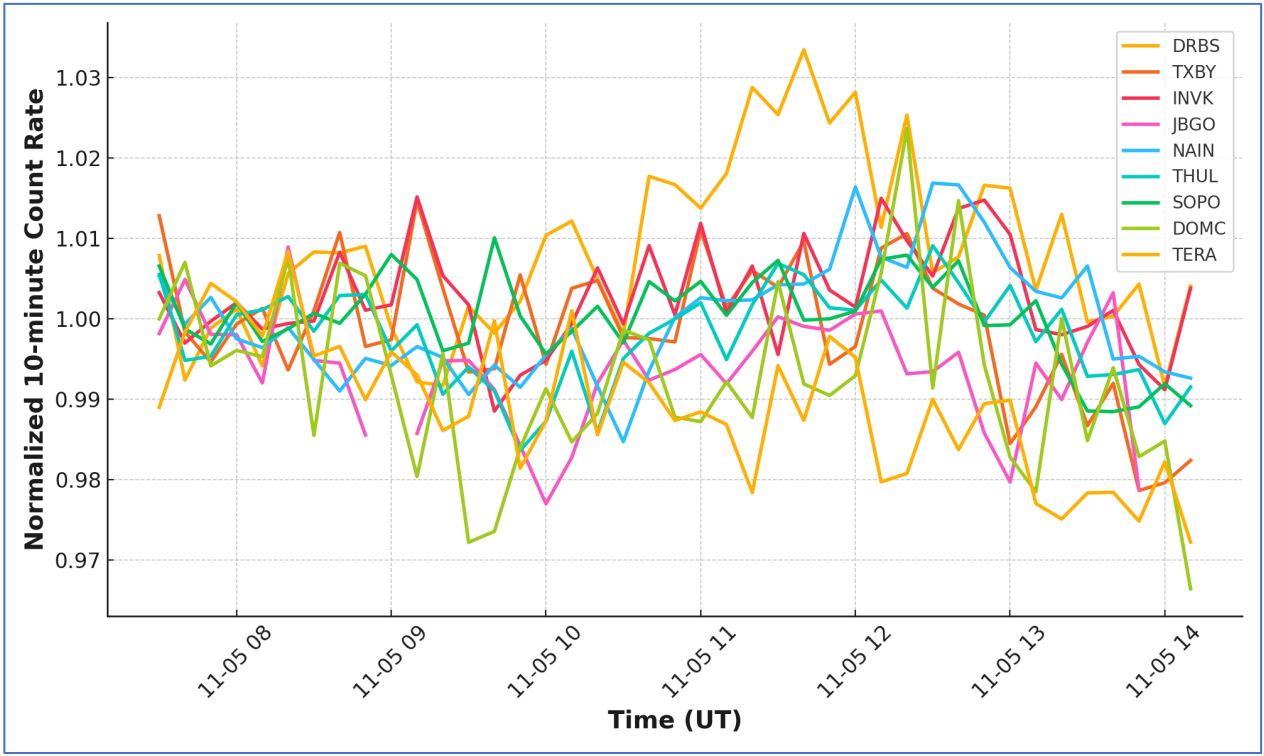

**Figure 1. Normalized 10-minute time series of the Antarctic, near-polar, and Belgium neutron monitors during the solar event of 5 November 2023**

In Fig. 2, we present correlated enhancements (1.1-2.7%) of the time series of NM count rates at mid-latitudes and high altitudes. The time series is measured by NMs prolonged across more than 5500 km. The enhancements are similar in shape, and the correlations are obvious.

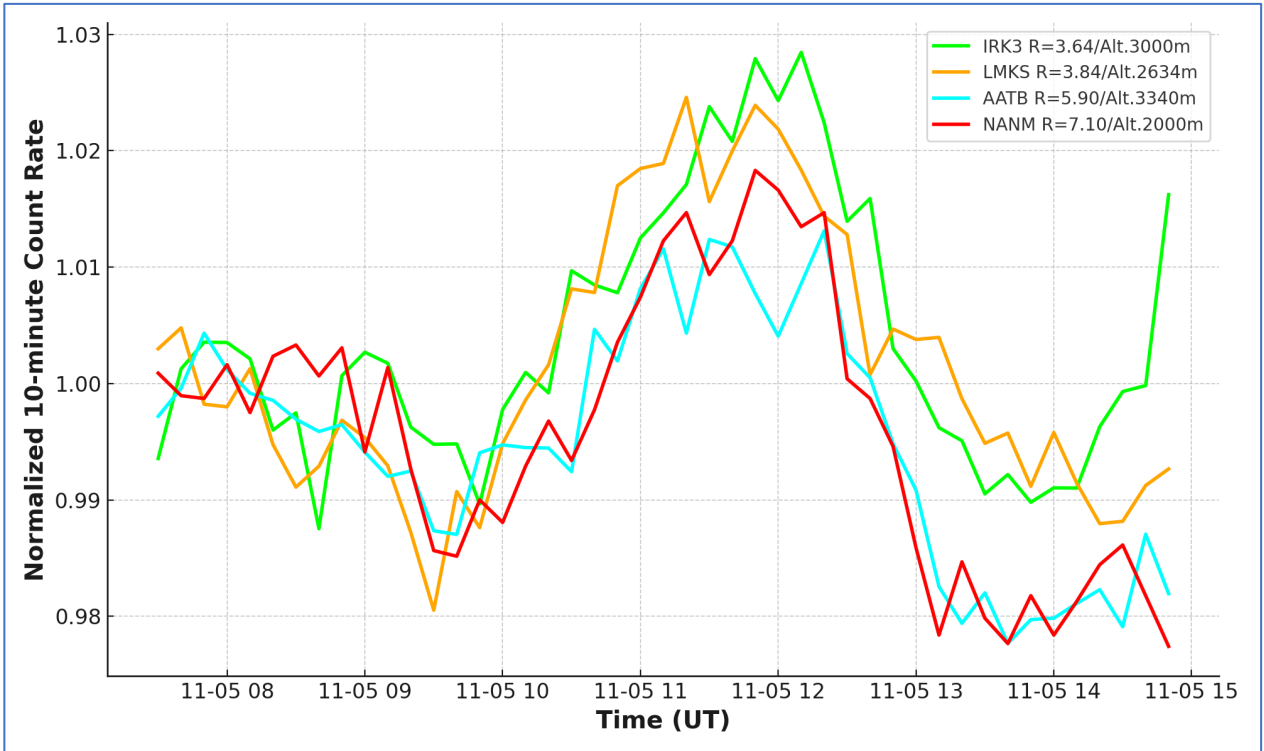

**Figure 2. Significant increase in the count rates of monitors situated at middle latitudes and high altitudes.**

**Figure 3 displays the time series of count rates from the SEVAN detectors at mountaintops and sea level.**

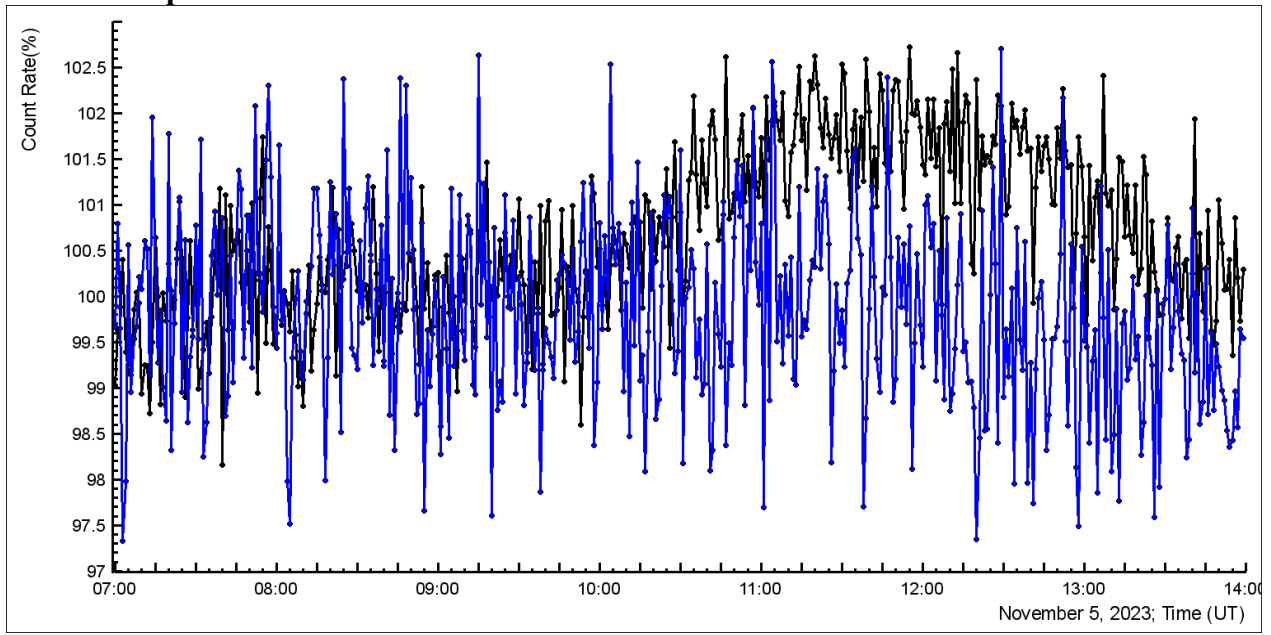

**Figure 3. 1-minute time series of count rates from SEVAN detectors at Lomnicky Stit (black, 2635 m elevation) and Hamburg (blue, 7 m elevation).**

Due to an atmospheric cutoff, the Hamburg SEVAN does not observe any enhancement, resulting in a missing ME. In contrast, the SEVAN at Lomnicky Stit Mountain exhibits a pronounced enhancement. Refer to Figs. 5 and 6 of Chilingarian et al.. 2024b.

In Fig. 4a, we compare the count rates of the SEVAN detectors at Aragats and Lomnicky Stit (for details on SEVAN detector operation, see the instrumentation section of Chilingarian et al., 2024c). The dynamic count rates of both detectors, located at distances of approximately 2000 km, are very similar, and the correlation coefficient shown in Fig. 4b is 0.6. The ME enhancement reaches 2-2.5%. The "100" coincidence of the SEVAN detectors (with a signal only present in the upper of three stacked scintillators) selects particles with energies exceeding approximately 5 MeV. The count rate on fair-weather of the Aragats SEVAN, situated at an altitude of 3200 m, is approximately 5% higher than that at Lomnicky Stit, which is at 2635m.

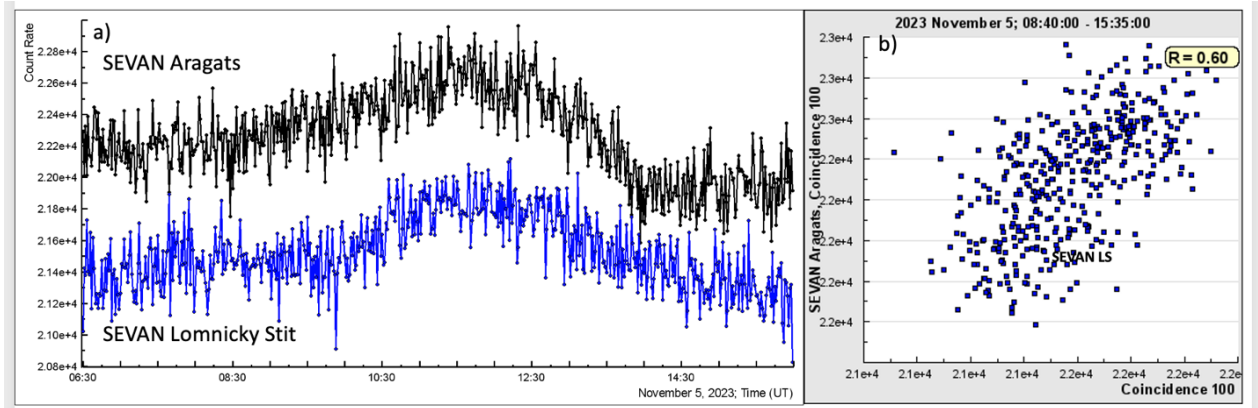

**Figure 4. a) 1-minute time series of the "100" coincidences from the SEVAN detector at Aragats and Lomnicky Štít; b) scatter plot of count rates for both SEVAN detectors.**

2. Comparison of the GLE event on May 11, 2024, and the ME event on 5 November, 2023

The geomagnetic storm classified as G5 took place on May 10-11, 2024, marking one of the most powerful storms in the past twenty-plus years (for an in-depth examination, refer to Hayakawa et al., 2024). This phenomenon commenced with notable FD and transitioned into GLE #74 during FD's recovery phase. Ground-based neutron monitors (Mavromichalaki et al., 2011) validated a GLE associated with 02 UTC on May 11, 2024, following an X.5.8 flare that peaked at 01:23 UTC. Figure 5 presents a detailed time series of the precise correlated counts from the 5 cm thick, one m² area SEVAN upper scintillators located on Mts. Aragats (40.25°N, 44.15°E, altitude 3200 m), Musala (42.1°N, 23.35°E, altitude 2930 m), and Lomnicky Stit (49.2°N, 20.22°E, altitude 2635 m).

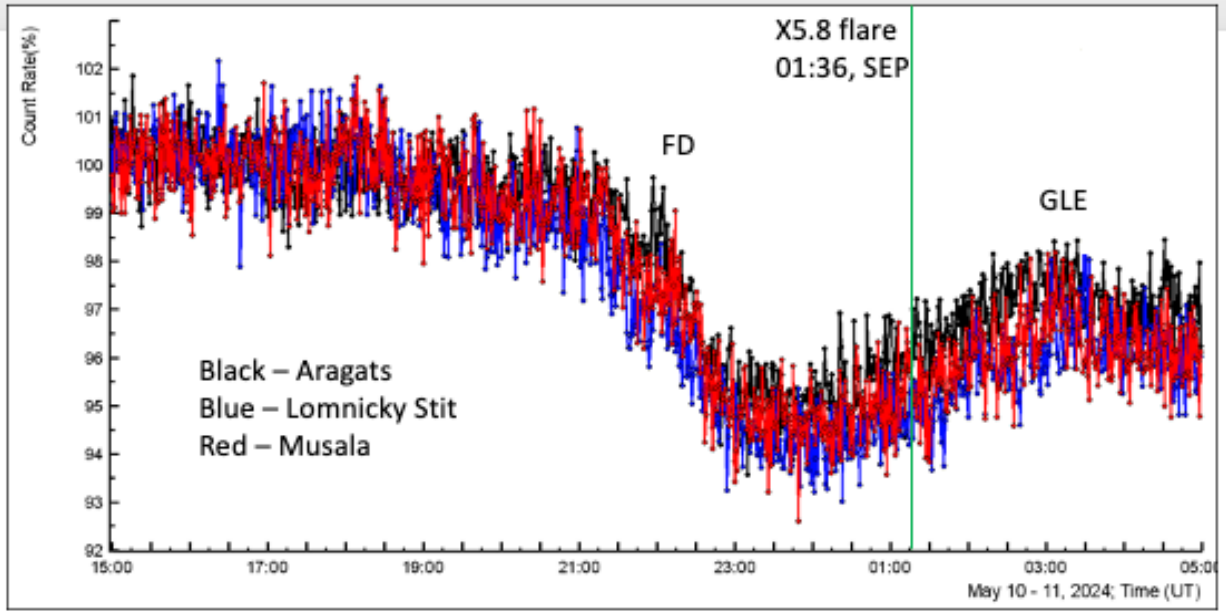

**Figure 5. Observation of FD and GLE by SEVAN network's Aragats, Lomnicky Stit, and Musala detectors**

Figure 6 illustrates the energy spectra of GLE and ME obtained from energy release histograms, measured using a scintillation spectrometer with a thickness of 20 cm and an area of 0.25 m². We present the energy spectra recorded at the peak minutes of fluxes. We subtract the background (CR) measured before the solar events from the count rates recorded during solar events (CR + ME and CR + GLE). For information on detector operation, see Chilingarian et al. (2024a) and the spectrum recovery methods detailed in Chilingarian et al. (2022). The maximum energy of GLE particles is tenfold larger than the maximum energy of particles registered during ME. The ≈10 GeV SEP entering the atmosphere produces secondaries (neutrons and muons) with energies above 100 MeV. The energies of ME particles are limited to 10 MeV, indicating low-energy primary protons (below 7.1 GeV on Aragats) contributing to this solar event. The GMS-induced weakening of geomagnetic shielding enables primary GCR with energies below the geomagnetic cutoff rigidity to penetrate the atmosphere and produce EASs. In contrast, the secondaries from SEP on May 11 (see details on the GLE event in Chilingarian et al., 2024c) extend beyond 100 MeV, as shown in Fig. 6b. As we demonstrate in Bostanjan et al. (2007) and Chilingarian (2009), the energies of solar protons can reach 20 GeV or more. Therefore, during GLE, SCR can generate secondaries with energies significantly larger than those during ME.

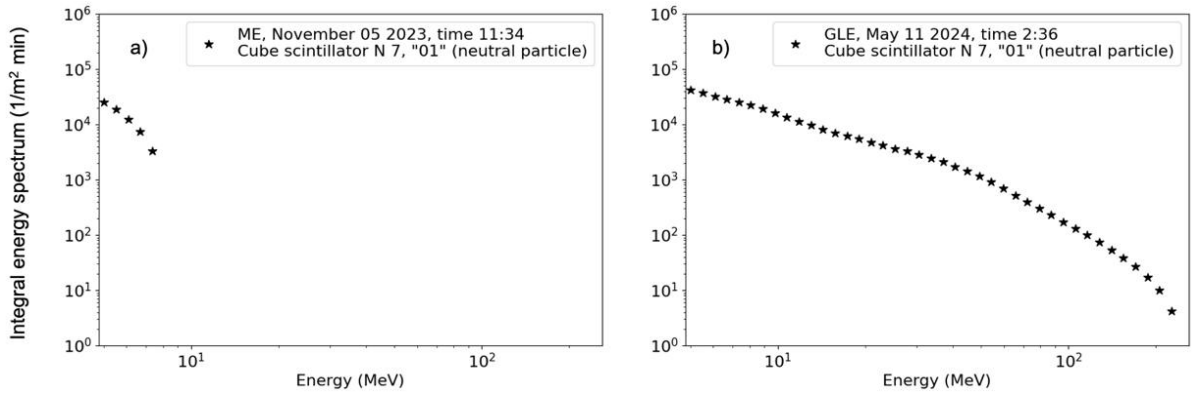

**Figure 6. a) the integral energy spectra of ME particles, b) the integral energy spectra of GLE particles measured by the spectrometer on Aragats.**

The GCRs causing ME on November 5, 2023, have lower energy than the minimal energies allowed by geomagnetic cutoff rigidity. Consequently, they produce smaller particle showers with secondary particles of lower energy. Thus, they reach mountain altitudes and sea level in far smaller amounts due to atmospheric rigidity. As a result, there was no flux enhancement at sea level (Figs. 1 and 3). The intensity of secondary electrons (Fig. 7a) and gamma rays (Fig. 7b) generated by 7 GeV protons (the Rc of Aragats is 7.1 GV) entering the atmosphere is significantly greater at high altitudes (Tibet, Aragats, black and red curves) than at lower altitudes (Nor Amberd, green curve) and sea level (blue curve in Fig. 7).

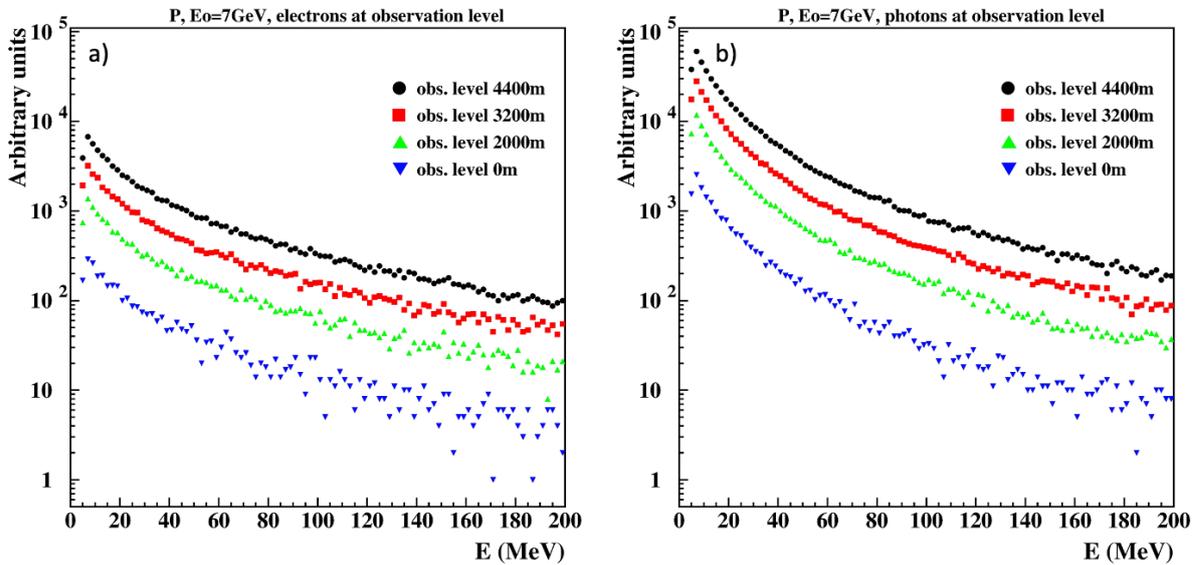

**Figure 7. Energy spectra of secondary electrons (a) and gamma rays (b) from the EASs originated by a primary proton with an energy of 7 GeV incident on the terrestrial atmosphere. The colored curves display energy spectra depending on the observation height level.**

To obtain the fluxes of secondary particles, we utilized version 7.75 of the CORSIKA code (Heck et al., 1998), which simulates extensive air showers (EASs) using the QGSII_UrQMD hadronic interaction models (Fesefeldt, 1985) and the EGS4 electromagnetic interaction model (Nelson et al., 1985). GeV protons enter the Earth's atmosphere vertically. We tracked shower electrons, positrons, and gamma rays until they reached energies of 5 MeV and muons until they reached 10 MeV, completing 1000000 simulation trials. Table 1 presents the average number of detected secondary particles produced by 7 GeV protons at various altitudes. On average, only 0.003 electrons from the initial 7 GeV proton reach sea level; at an altitude of 3200 m (Aragats station), this number rises significantly to an average of 0.04 electrons, over ten times higher. Thus, as illustrated in Figs 1-3, muons are effectively recorded at high altitudes but are not detected by instruments located at sea level.

In contrast, during a Ground Level Enhancement (GLE) event, solar protons generate secondary particles with energies that exceed the cutoff rigidity. These high-energy protons produce secondary particles that can reach sea level, enabling ground-based detectors to observe GLE. As a result, the energies of secondary particles during GLE events are considerably higher than those noted during muon events.

**Table 1. Average number of detected secondary particles per primary proton at various observation levels**

| Obs. Level (m) | µ+&µ- | e+&e- | γ |
|---|---|---|---|
| 0 | 0.011 | 0.003 | 0.018 |
| 2000 | 0.032 | 0.016 | 0.090 |
| 3200 | 0.054 | 0.040 | 0.218 |
| 4400 | 0.084 | 0.086 | 0.468 |

Consequently, the energy spectra of secondary particles (GLE particles) can provide insights into the spectra of solar energetic particles (SEP spectra). Since GLE spectra can be monitored online on a one-minute timescale, this allows for alerts regarding dangerous "hard" SEPs, which have spectral indices of 4-5 at GeV energies (Chilingarian and Reymers, 2007). Monitoring the GLE energy spectra, as shown in Fig. 8, enables us to track shifts in the GLE spectrum. In Fig. 8, we observe that the spectra decayed more quickly at both the beginning and end of GLE (red and rose asterisks). At the same time, the maximum energy of neutral particles extends above 100 MeV at the GLE maximum flux (black asterisks). The relationship between SEP and GLE spectra can be established through multiple simulations involving various SEP parameters, as clarified in Zazyan and Chilingarian (2009). Figure 10 shows the simulated neutron energy spectra at 3200 m corresponding to SEP energies from 7 to 20 GeV. We performed 1 million simulations for each SEP energy, with the neutron cutoff energy set at 50 MeV.

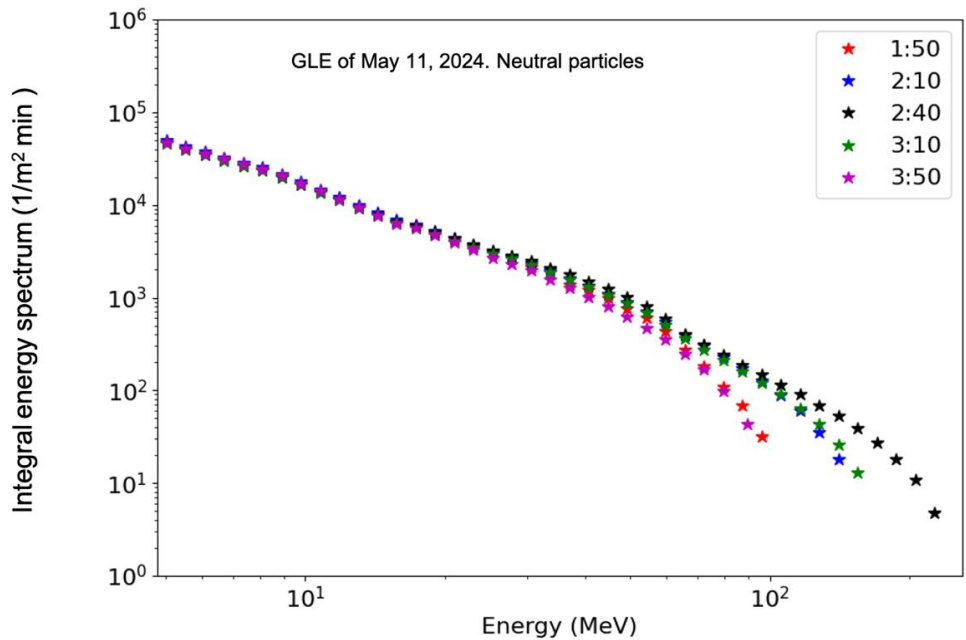

**Figure 8. GLE energy spectra followed from the beginning (red) to the end (rose).**

Figure 9 shows a strong dependence of the energy spectra on primary proton energy. At 300 MeV, we expect only 2 neutrons from ten thousand SEPs with energy 7 GeV and approximately 150 from 20 GeV SEPs. Therefore, when extended to high energies, the GLE spectrum indicates a hard SEP event, which is dangerous for satellite electronics.

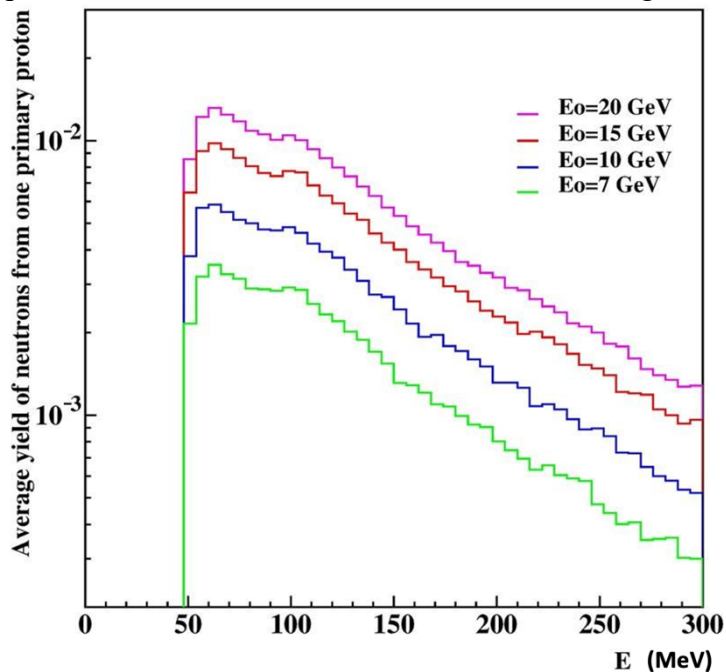

**Figure 9. The average neutron yield from vertically incident solar energetic protons of various energies reaches 3200 m.**

**Conclusions**

The surprising increase in solar activity during the fall of 2023's 25th solar cycle, after a relatively tranquil 24th cycle, suggests we are approaching the solar maximum of the current cycle. The complex interplay between interplanetary and geomagnetic fields has various effects, from harm to satellite electronics to stunning auroras. Given these developments, comprehending solar activity's impact on Earth's environment is becoming more crucial. Cosmic rays are important indicators, providing essential information about these complex processes. Networks of particle detectors that consistently track cosmic ray flux on the Earth's surface yield valuable insights, augmenting data gathered by space-borne detectors managed by NOAA, NASA, and ESA

We present neutron energy spectra related to the magnetospheric effect observed on November 5, 2023, and the GLE on May 11, 2024. The significant difference between the two highlights the importance of measuring energy spectra for identifying solar energetic events.

- ME events are driven by low-energy galactic cosmic rays below the geomagnetic cutoff rigidity, and they register revealingly at middle and low latitudes and high altitudes where atmospheric rigidity is lower.
- GLE events are initiated by higher-energy solar cosmic rays that can penetrate Earth's magnetosphere and atmosphere at all latitudes and heights.

The SEVAN network enables the measurement of energy spectra of secondary particles associated with solar events (SEP and ME). This allows for examining the energy spectra of solar energetic protons (SEP) by simulating showers initiated by solar protons of varying energies. By identifying correlations between secondary neutrons and muons (MeV energies) and SEP of GeV energies, particle spectrometers will help recognize SEP events with hard spectra, which pose significant risks to satellite electronics and surface operations. This capability can be extremely valuable for issuing alerts regarding impending hazardous SEP events, as energy spectra can be retrieved online.

In addition, we refine the definition of the Magnetospheric Effect (ME). Our classification is based on the following observations:

1. Antarctic and near-polar NMs did not register flux enhancement, while middle latitude NMs and SEVAN detectors coherently registered count rate enhancements across distances of 5,500 km.
2. Due to atmospheric cutoff rigidity, NMs and SEVAN detectors located on mountain tops at middle latitudes demonstrated flux enhancement, while ones at sea level did not.
3. The energy spectra of the shower particles measured during ME are significantly different from those of other sources of flux enhancement, making this criterion essential for identifying ME.

Another model is supported in the Gil et al. (2024) paper. According to this model, the increase in CR flux on 5 November 2023 was due to 'focusing of GCRs in strongly disturbed heliospheric conditions', rather than a decrease in the geomagnetic cutoff rigidity. However, we stand by our interpretation of the event as a global magnetospheric effect. We do not agree that "effect was observed at high-latitude NMs, including the South Pole, as well as at stations at sea level". Fig. 1 of our paper and Fig. 1 of Gil's paper demonstrate that only the DRBS monitor reliably shows apparent enhancement. However, it is not in Antarctica but in the Geophysical Center of Dourbes (Belgium).

Furthermore, the energy spectra of secondary particles presented in Fig. 7 prove that the parent GeV protons have low energies, below 7.1 GeV (see also Fig. 8 and Table 1). We demonstrate a strong dependence of the energy spectra of secondaries on primary proton energy. We measure the energy spectra of GLE neutrons, shown in Fig. 9. Fig. 10 demonstrates the yield of secondaries and their energies from parent protons ranging from 7 to 20 GeV, proving the observed striking difference between the ME and GLE events.

We are confident that our arguments are strong enough to confirm the ME origin of the November 2023 event. Certainly, other groups may support alternative models. However, competing models should be published equally, as alternative hypotheses have the right to be presented to the community.

**Acknowledgments**

We thank our Neutron Monitor Database (NMDB) collaboration colleagues for their insightful discussions. Additionally, we recognize the NMDB database (www.nmdb.eu), created under the European Union's FP7 program (contract No. 213007), for granting access to neutron monitor data.

Data availability statement: The data supporting the findings of this study are available at the URL/DOI: http://adei.crd.yerphi.am/